\documentclass[3p,times,twocolumn]{elsarticle}

\usepackage{ecrc}

\volume{00}

\firstpage{1}

\journalname{Nuclear and Particle Physics Proceedings}

\runauth{Mirta Duman\v{c}i\'{c}}

\jid{nppp}

\jnltitlelogo{Nuclear and Particle Physics Proceedings}

\usepackage{amssymb}
\usepackage{amsmath}

\usepackage{lineno}

\usepackage[figuresright]{rotating}
\usepackage{xspace}

\newcommand*{\pPb}{\ensuremath{p}+Pb\xspace}
\newcommand*{\PbPb}{Pb+Pb\xspace}
\newcommand*{\Zboson}{\ensuremath{Z}\xspace}
\newcommand*{\Wboson}{\ensuremath{W}\xspace}

\newcommand*{\pp}{\ensuremath{pp}\xspace}

\newcommand*{\sqn}{\ensuremath{\sqrt{s_{_{\mathrm{NN}}}}}\xspace}
\newcommand*{\sqs}{\ensuremath{\sqrt{s}}\xspace}

\newcommand{\yZ}{\mbox{$y^{Z}$}\xspace}
\newcommand*{\TeV}{\mbox{TeV}\xspace}

\newcommand*{\Tppb}{\ensuremath{T_{p\mathrm{Pb}}}\xspace}
\newcommand*{\avgTppb}{\ensuremath{\langle\Tppb\rangle}\xspace}

\newcommand*{\RpPb}{\ensuremath{R_{p\mathrm{Pb}}}\xspace}

\newcommand*{\ipb}{\mbox{pb$^{-1}$}}
\newcommand*{\inb}{\mbox{nb$^{-1}$}}

\begin{document}


\begin{frontmatter}



\dochead{}

\title{\textit{W} and \textit{Z} boson production in 5.02 TeV \textit{pp} and \textit{p}+Pb collisions with the ATLAS detector}


\author{Mirta Duman\v{c}i\'{c}}
\author{on behalf of the ATLAS Collaboration}

\address{Department of Particle Physics and Astrophysics, Weizmann Institute of Science}

\begin{abstract}

This proceeding reports the ATLAS results of measuring vector boson production in \pp and \pPb collisions at the center-of-mass energy of 5.02~TeV per nucleon. \Zboson bosons are reconstructed via leptonic decays and results are presented as a function of $Z$-boson rapidity. \Wboson bosons are identified by measuring leptons coming from \Wboson decays and the results are presented as a function of lepton pseudorapidity. In \pp collisions \Zboson boson cross section in $|y_{Z}|<2.5$ is 590$~\pm$~9 (stat.)~$\pm$~12 (syst.)~$\pm$~32 (lumi.) pb. In \pPb system kinematic distributions of \Wboson and \Zboson bosons are compared to models based on NLO and NNLO QCD calculations using different PDF sets. Using PDF modification shows better agreement of the model and the data. Measurements done in different centrality intervals show that the PDF modification in \pPb may have centrality dependence. 

\end{abstract}


\end{frontmatter}


Experimental study of electroweak (EW) bosons in relativistic heavy ion (HI) collisions is an integral part of the physics program carried out by all four detectors taking data at the Large Hadron Collider (LHC). The results of these programs have demonstrated several important phenomena. Since EW bosons and their leptonic decay products do not interact strongly with the hot and dense matter created in the HI collision, their production rates are expected to be sensitive to the effective overlap area of colliding nuclear matter. This has been confirmed in measurements performed by the ATLAS and CMS experiments with \Zboson and \Wboson bosons decaying leptonically or semileptonically where it has been shown that the production rate of non-strongly interacting particles in \PbPb collisions scales with the nuclear thickness function~\cite{Aad:2012ew,Aad:2014bha,Aad:2015lcb,Chatrchyan:2011ua,Chatrchyan:2012nt}. \par
EW bosons are also used as an outstanding tool to study nuclear modifications to parton distribution functions (PDF). These effects include nuclear shadowing, anti-shadowing and the EMC effect. In particular, rapidity distributions of \Zboson and \Wboson bosons determined by the Bjorken $x$ of the interacting partons are sensitive to the presence of nuclear modification. The existing \PbPb measurements due to their limited precision cannot exclude the presence of the nuclear modification~\cite{Aad:2012ew,Aad:2014bha,Aad:2015lcb,Chatrchyan:2011ua,Chatrchyan:2012nt}. \par 
Study of asymmetric collisions systems, such as proton-lead (\pPb) can be used to differentiate between initial and final state effects in HI collisions. The results using 2013 \pPb data at the center-of-mass energy, \sqn ~=~5.02~\TeV on EW boson production have been published by all LHC experiments ~\cite{Aad:2015gta,Khachatryan:2015pzs,Khachatryan:2015hha,Alice:2016wka,Aaij:2014pvu}. 
Using preliminary result~\cite{me:2016} on \Zboson boson production in \pp collisions at the same center-of-mass energy as in \pPb it becomes possible to study modifications observed in that system with respect to a physics measurement rather than to calculated prediction. \par






The \pPb data obtained by the ATLAS experiment at \sqn=5.02~TeV corresponding to integrated luminosity of 29~\inb \xspace has been used to measure \Zboson boson production in muon and electron decay channels~\cite{Aad:2015gta}. Events for analysis in the electron channel were selected by high-level trigger requiring an electron with at least 15 GeV transverse momentum that also passes loose identification criteria. Similarly, events in the muon channel were selected with the high-level trigger requiring a muon with transverse momentum of at least 8 GeV. In total 1647 (2032) \Zboson boson candidates were reconstructed in electron (muon) decay channel. In particular, electrons within the range $2.5<|\eta|<4.9$ are reconstructed based on the energy deposited in the forward calorimeter that allows for the reconstruction of \Zboson boson candidates up to $|y^{Z}|<3.5$. This yields additional 264 \Zboson boson candidates. Total measured \Zboson boson fiducial cross section of 139.8~$\pm$~4.8 (stat.)~$\pm$~6.2 (syst.)~$\pm$~3.8 (lumi.)~nb is found to be slightly higher compared to model predictions based on perturbative QCD (pQCD) calculations that include CT10 PDF set~\cite{CT10ref}. \par



Preliminary measurement of the \Zboson boson cross section has been obtained from the \pp data sample corresponding to the integrated luminosity of 24.7 $\pm$ 1.3 ~\ipb\  at  \sqs~=~5.02~\TeV. Events for the analysis have been selected with the high-level trigger requiring a muon with transverse momentum of at least 14 GeV. In total 7293 \Zboson boson candidates passed all the analysis selection. 
The \Zboson boson fiducial cross section is measured to be 590~$\pm$~9~(stat.)~$\pm$~12~(syst.)~$\pm$~32~(lumi.)~pb. Model with CT10 PDF set predicts a significantly lower cross section of 537~pb. The NNLO prediction using the CT14 PDF set~\cite{CT14ref} and calculated using a version of DYNNLO 1.5~\cite{dyn1,dyn2}  yields a cross section of $573.77^{+13.94}_{-15.96}$~pb which agrees well with the measurement within its uncertainties.

\begin{figure}[ht!]
\centering
\includegraphics[width=0.42 \textwidth]{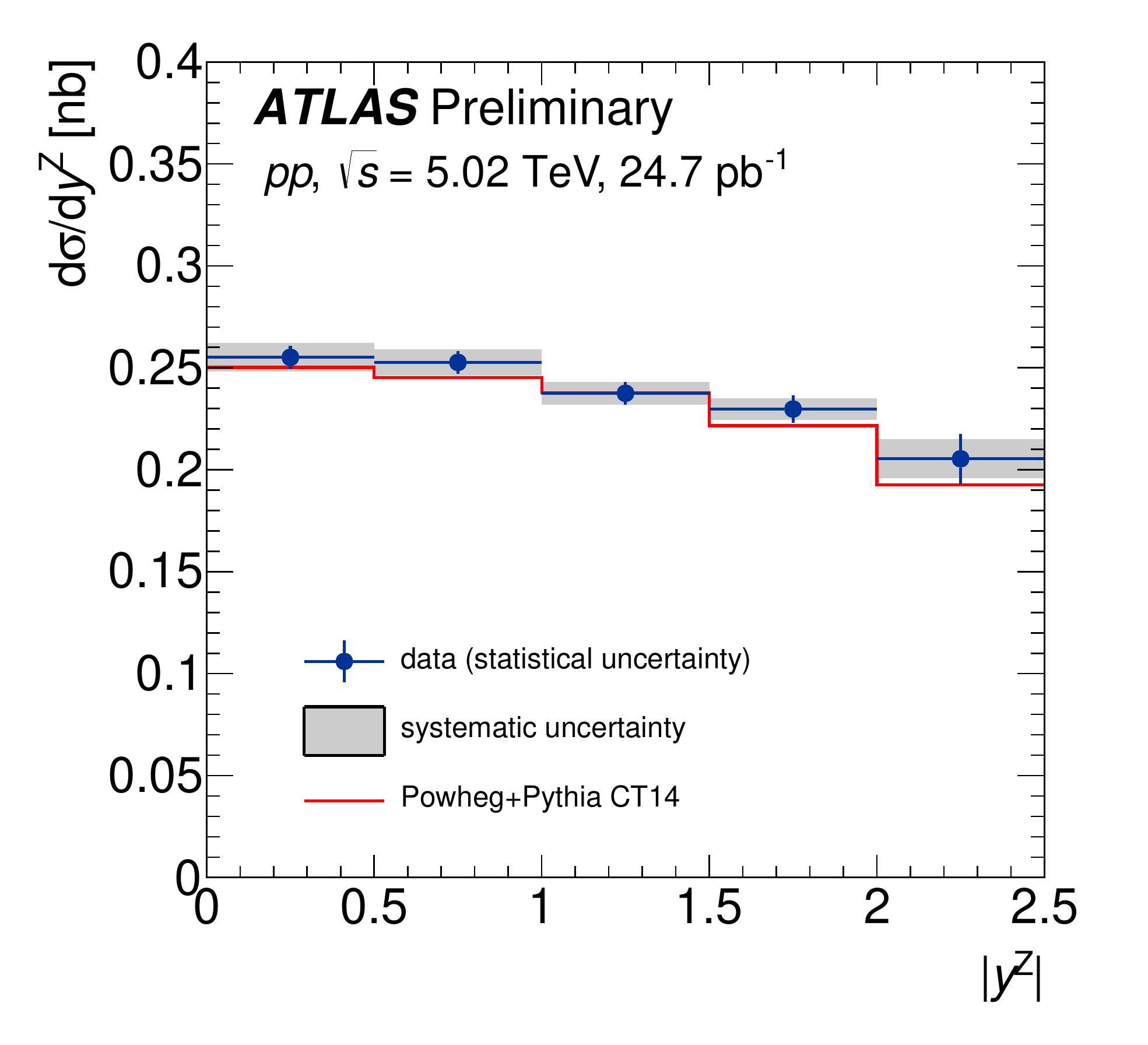}
\caption{The measured rapidity differential cross section in \pp data.  The data is compared to the simulated MC generated with the CT10 PDF at NLO, scaled to the integrated cross section calculated with DYNNLO at NNLO using the CT14 PDF set~\cite{me:2016}.}
\label{fig:Xsection}
\end{figure}


In addition to the integrated cross section, \Zboson analyses in \pPb and \pp datasets include measurements of the \yZ\ differential cross section. Figure~\ref{fig:Xsection} shows differential cross section $\textrm{d}\sigma/\textrm{d}y^{Z}$ measured in \pp collisions.  Because of the symmetry in \yZ, the data is shown in bins of $|\yZ|$.  The data is compared to the simulated MC generated with the CT10 PDF at NLO, scaled to the integrated cross section calculated at NNLO using the CT14 PDF set.  The scaled simulation is in good agreement with the data. \par

\begin{figure}[ht!]
\centering
\includegraphics[width=0.4 \textwidth]{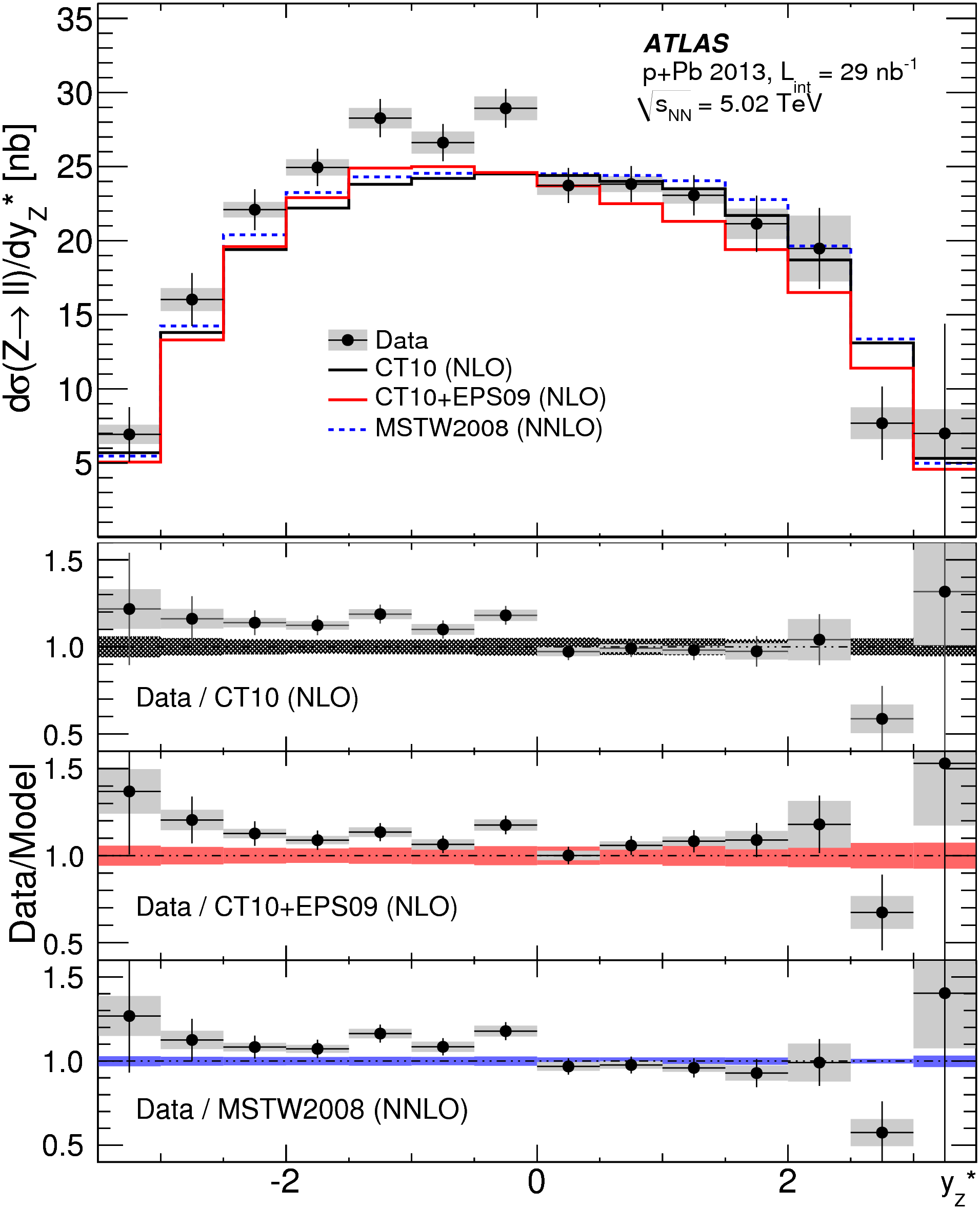}
\caption{ (a) The $d\sigma/dy^{Z}$ distribution from \Zboson boson decays measured in \pPb data, shown along with several model calculations in the upper panel. The bars indicate statistical uncertainty and the shaded boxes systematic uncertainty on the data; the uncertainties on the model calculations are not shown. (b-d) Ratios of the data to the models. The uncertainties of the model calculations (scale and PDF uncertainties added in quadrature) are shown as bands around unity in each panel. An additional 2.7\% luminosity uncertainty on the cross section is not shown~\cite{Aad:2015gta}.}
\label{fig:pPbZ}
\end{figure}


The rapidity differential cross section measured in \pPb data is presented in Figure~\ref{fig:pPbZ} and compared to model calculations. The data shows a strong asymmetry about $y^{Z}=0$ compared to the model prediction with CT10 PDF set and is better described by the models containing nuclear modification such as EPS09.  

\begin{figure}[ht!]
\centering
\includegraphics[width=0.42 \textwidth]{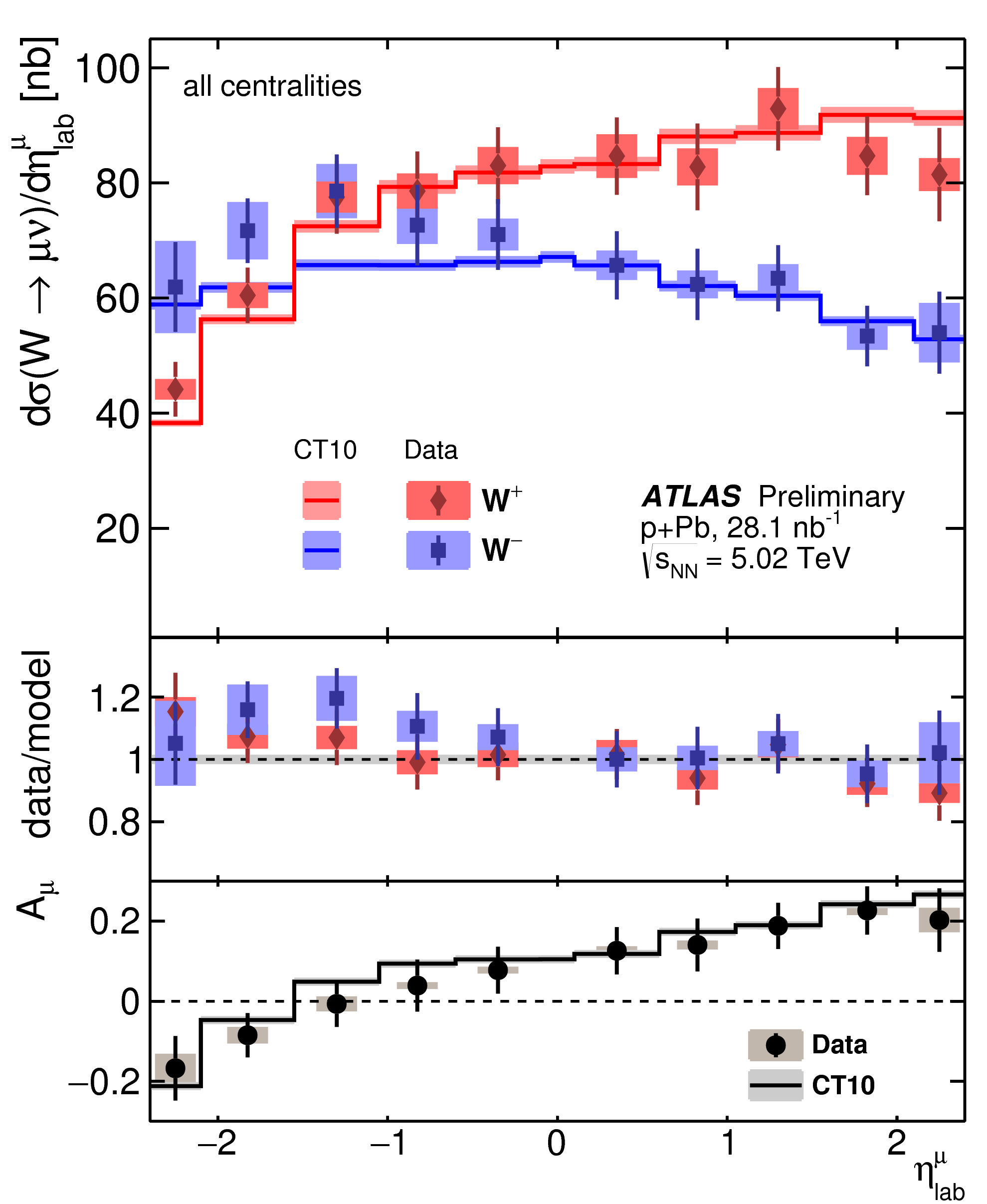}
\caption{The upper panel shows the cross section of the $W^{+}$ and $W^{-}$ production measured in \pPb data as a function of the lepton pseudorapidity compared to model prediction based on CT10 PDF set. The middle panel shows the data-to-model ratios for and the lower panel shows the lepton charge asymmetry compared to the same model~\cite{Markus:2015}.}
\label{fig:WpPb}
\end{figure}

In the same \pPb dataset the $\Wboson \rightarrow \mu\nu$ production and the dependence of the cross section on the pseudorapidity of the muons has also been studied~\cite{Markus:2015}. Prediction based on pQCD calculations reproduce the data well, except for the $W^{-}$ boson in the lead-going direction where there is an excess above the model, which is consistent with the similar observation in the \Zboson boson measurement. In order to study the difference in production of $W^{+}$ and $W^{-}$ bosons, the observable called lepton charge asymmetry  $A_{\mu}(\eta_{\mu})$ has also been measured. Measurement shows deviation from the CT10 prediction on the lead-going side. Upper panel of Figure~\ref{fig:WpPb} shows $W^{+}$ and $W^{-}$ cross section for all centrality classes of events as a function of the lepton pseudorapidity compared to model prediction based on CT10 PDF set. Middle panel shows the ratio to the model for different charges and the lower panel shows the lepton charge asymmetry compared to the model.  
\par

\begin{figure}[h!]
\centering
\includegraphics[width=0.45 \textwidth]{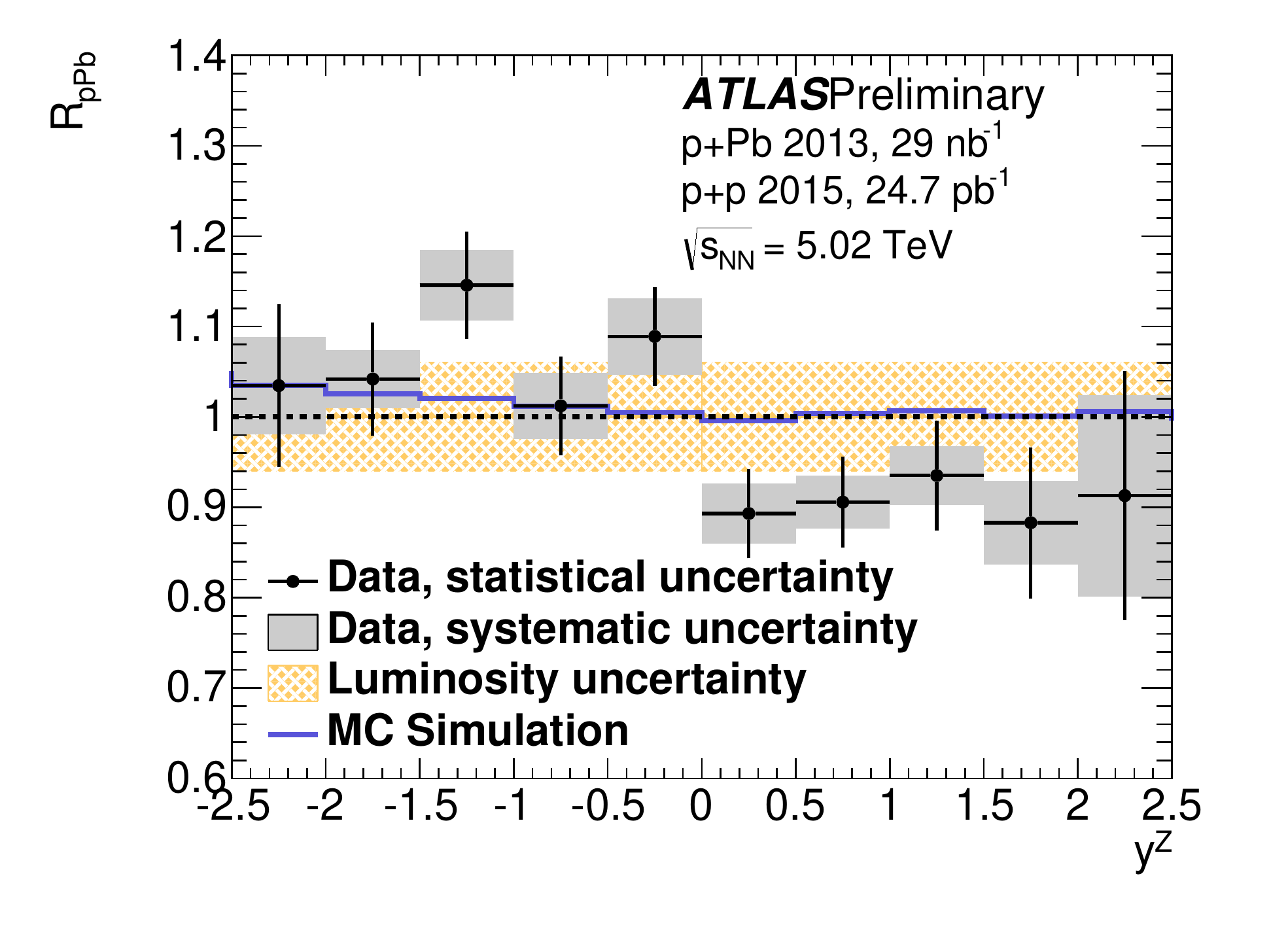}
\caption{The nuclear modification factor as a function of \Zboson boson rapidity. The uncertainties on each point include statistical and systematic uncertainties from \pPb and \pp measurements.  The band around unity represents the uncertainty of the luminosities. The blue line shows an expected \RpPb based on simulation~\cite{me:2016}.}
\label{fig:Rpa}
\end{figure}

The \Zboson boson cross section measured in \pp collisions may be used as a baseline to study nuclear modifications which may be present in the previously measured cross section in \pPb collisions~\cite{Aad:2015gta}. 
The nuclear modification factor, \RpPb, is defined as the ratio of the cross sections measured in \pPb and \pp systems. The exact definition of \RpPb can be found in~\cite{me:2016}. Figure~\ref{fig:Rpa} presents the nuclear modification factor as a function of \yZ for all centralities of events in \pPb data. It shows enhancement in the lead-going direction and suppression in the proton-going direction. The blue curve which describes the expected contribution from the isospin effect is not sufficient to describe the shape of the measured distribution. This difference can be explained by nuclear PDF modification present inside the lead nucleus and effects observed in \pPb measurements as discussed above.  \par

\begin{figure}[ht!]
\centering
\includegraphics[width=0.42 \textwidth]{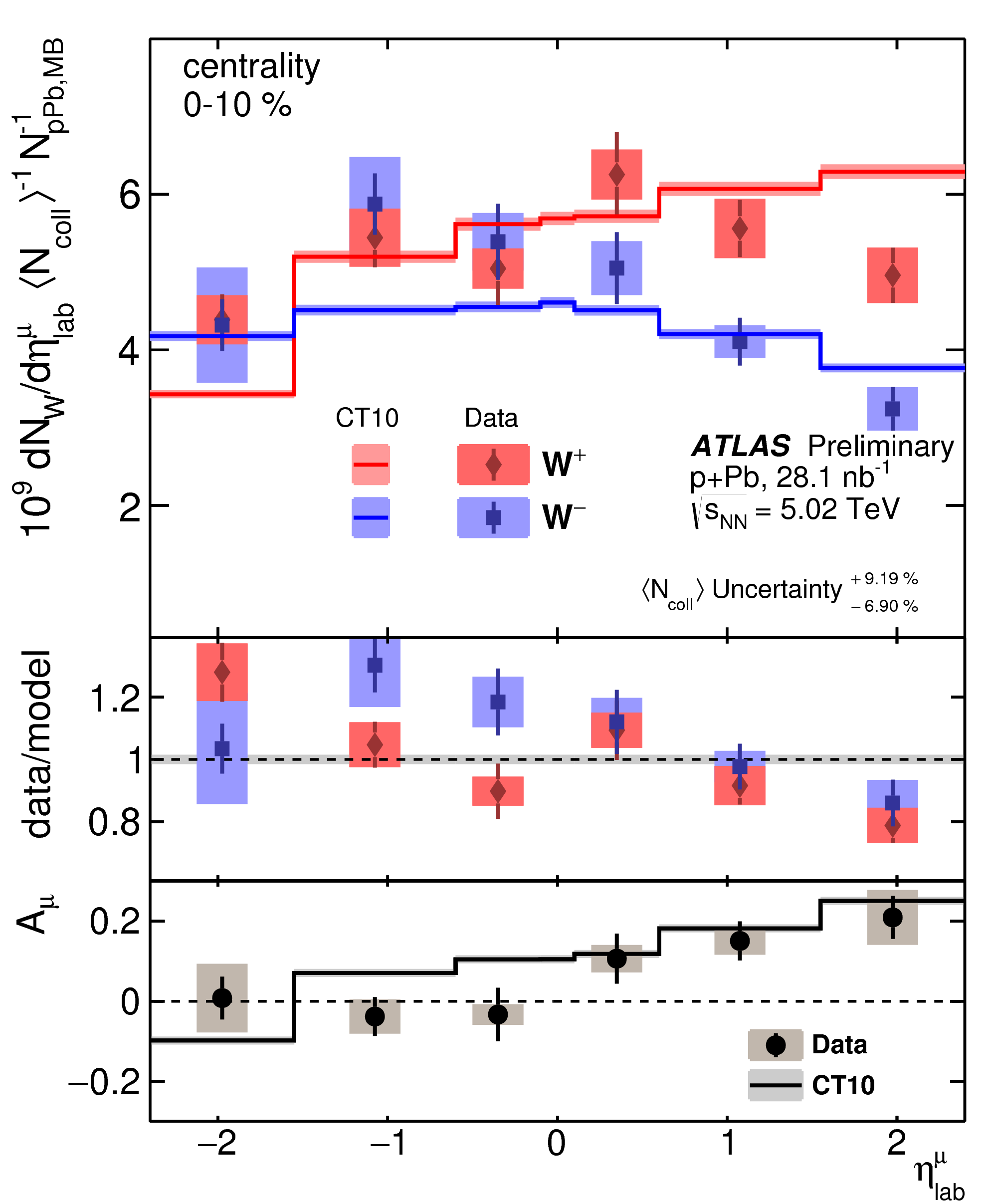}
\caption{The upper panel shows the cross section of the $W^{+}$ and $W^{-}$ production as a function of the lepton pseudorapidity compared to model prediction based on CT10 PDF set measured in 0-10\% centrality interval of the \pPb data. The middle panel shows the data-to-model ratios for and the lower panel shows the lepton charge asymmetry compared to the same model~\cite{Markus:2015}.}
\label{fig:WpPb_010}
\end{figure}

Measurements suggest that the nuclear modification factor has dependance on collision centrality~\cite{Aad:2015gta, Markus:2015}. Figure~\ref{fig:WpPb_010} shows the $W^{+}$ and $W^{-}$ boson differential yields in lepton pseudorapidity (upper panel) together with ratio to the model prediction (middle panel) and lepton charge asymmetry (bottom panel)  for events in the 0-10\% centrality class. There appears to be a dependence of the shape of the pseudorapidity distributions of both positively and negatively charged muons from \Wboson bosons on centrality. The data shown in the middle panel suggests the presence of a slope in most central collisions. The asymmetry shown in the lower panel also shows deviation from the prediction. This is similar to the trend observed in the \Zboson boson measurement as discussed in~\cite{Aad:2015gta}.

\begin{figure}[ht!] 
\centering
\includegraphics[width=0.45 \textwidth]{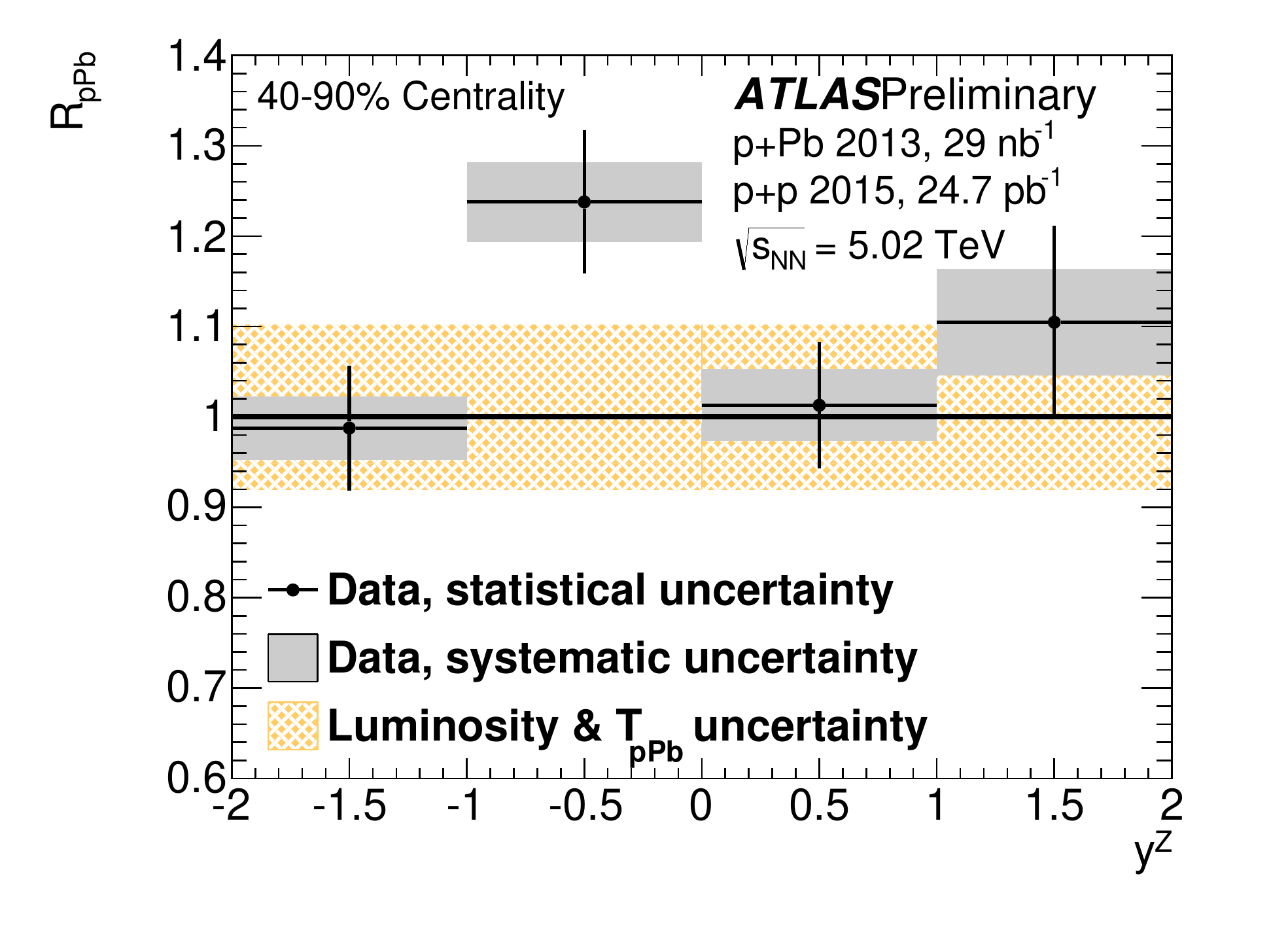}
\includegraphics[width=0.45 \textwidth]{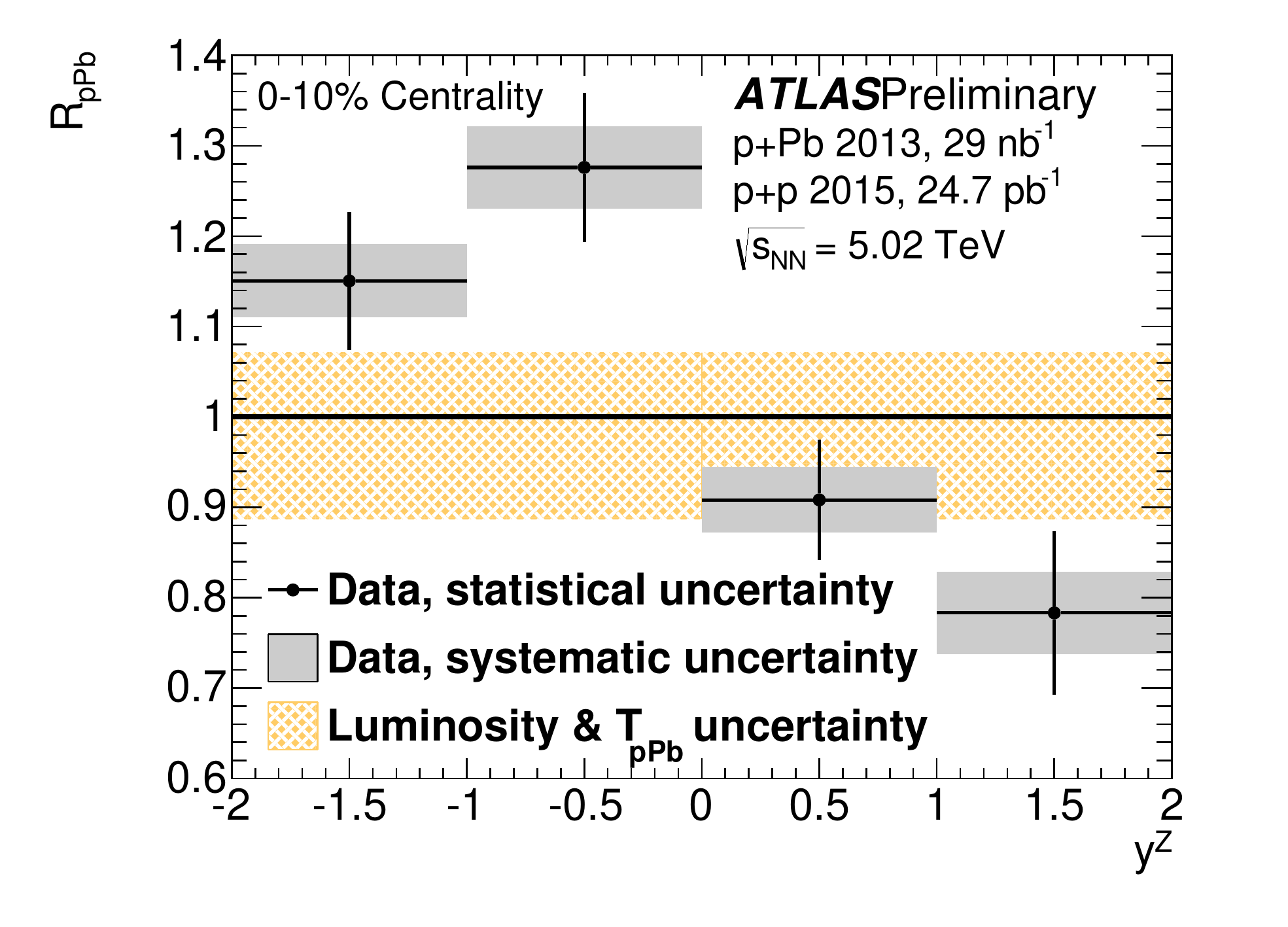}
\caption{The nuclear modification factor in most peripheral (40-90\%) and most central events (0-10\%). The uncertainties on each point include statistical and systematic uncertainties from \pPb and \pp measurements.  The band around unity represents the uncertainty of \avgTppb and luminosity~\cite{me:2016}.}
\label{fig:Rpa_centrality}
\end{figure}

Nuclear modification factor as a function of \yZ was measured for three different centrality classes of \pPb events. Results suggest that the asymmetry observed in the rapidity distribution of the \RpPb is more pronounced for more central collisions as shown in Figure~\ref{fig:Rpa_centrality}.  To quantify the change in asymmetry in different centrality classes, each $\RpPb^{\text{cent}}$ distribution is fitted to a linear function.  The resultant slopes for 40-90\%, 10-40\% and 0-10\% centrality bins are $0.02 \pm 0.04$, $-0.05 \pm 0.03$ and $-0.14 \pm 0.04$, respectively.  The uncertainties come from the fitting and are dominated by the uncertainties of the \pPb data. \par


\section*{Acknowledgements}
This research is supported by the Israel Science Foundation (grant 1065/15) and by the MINERVA Stiftung with the funds from the BMBF of the Federal Republic of Germany.




\bibliographystyle{elsarticle-num}

\bibliography{Z_2015_paper,ATLAS}







\end{document}